\documentstyle[epsfig,graphics]{article}
\def\fm{\hbox{$.\!\!^{\mathrm m}$}}
\def\fd{\hbox{$.\!\!^{\mathrm d}$}}

\def\fg{\hbox{$^\circ$}}
\begin{document}
\begin{center}
{\noindent\LARGE\bf HP Lyr -- possibly the hottest RV Tau type
object}\\[0.5cm]
        {\bf\large Dariusz Graczyk$\,^{1,4}\!\!$,
        Maciej Miko{\l}ajewski$\,^1\!$,
        Laurits Leedj{\"a}rv$\,^2\!$,
        Sylwia M.~Fr\c{a}ckowiak$\,^1\!$, 
        J\c{e}drzej P.~Osiwa{\l}a$\,^1\!$, 
        Alar Puss$^{2,3}\!\!$,
        and~Toma~Tomov$\,^1\!$}\\[0.2cm]
           {\small $^{1}$Centre for Astronomy, Nicolaus Copernicus University, 
           Gagarina 11, 87-100 Toru{\'n}, Poland\\
           e-mail: (DG) weganin@astri.uni.torun.pl, (MM) mamiko@astri.uni.torun.pl\\ 
           $^{2}$Tartu Observatory, 61602 T{\~o}ravere, Estonia, e-mail:~leed@aai.ee\\
           $^{3}$Department of Physics, Tartu University, 4 T{\"a}he Street, 51010
           Tartu, Estonia\\
           $^{4}$Institute of Astronomy, Lubuska 2, 65-265 Zielona G{\'o}ra, Poland}\\[0.7cm]
           
 ABSTRACT\\[0.2cm]
\end{center} 

{\small We report Johnson's $UBVRI$ photometric and optical spectroscopic 
observations of a long period variable HP Lyr which up to now
has been considered an eclipsing binary with a period of 140 days. 
The spectral type changes continually from A2-3 at maxima to A7-F2 at minima.
We propose that the brightness
changes are caused by the pulsation of the star with two periods: $P_1=69.35$, and 
$P_2=2\times P_1=138.7$ days. These periods decreased by more than 1\% between 1960 and 1980.
The spectral luminosity class determination gives an A type supergiant Iab.
HP Lyr is also the optical counterpart of the infrared 
source IRAS 19199+3950. Relatively high galactic latitude 
($b=+11^{o}\!\!.7$) and high radial velocity ($-$113 km/s) indicate 
that HP Lyr is an evolved, most likely post-AGB star. 
All presented features argue that this star is an RV Tau type object. }\\
  
\section{Introduction}

The star HP Lyr is a poorly known long-period variable 
discovered by Morgenroth (1935). 
The photographic photometry showed the light curve of a semiregular
type with the amplitude of brightness variations $\Delta m_{pg}=0\fm 5$ 
and the period of 70.4 days (Sandig 1939). 
The brightness at maximum was 10\fm 5 in $m_{pg}$ and no secular changes of
it were reported.
Wenzel (1960) determined the spectral type of the
variable as A6 and proposed that HP Lyr be an eclipsing binary  
consisting of a pair of A6 stars orbiting in a circular orbit with 
the period of $\sim$ 140 days. He classified the light curve as a 
$\beta$ Lyrae type with both minima of similar depth.

\section{Observations}
%

\begin{figure}
\begin{minipage}[th]{\linewidth}
\centering\epsfig{file=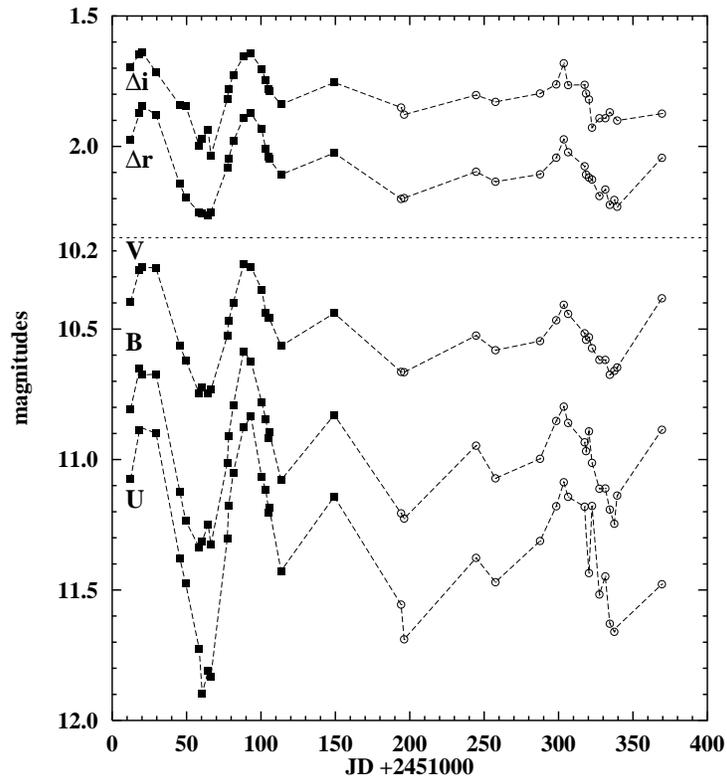,width=0.8\linewidth}
  \caption{\small Johnson's $UBV$ and instrumental $\Delta r$, $\Delta i$ light curves of HP Lyr in 1998/99 
obtained in Piwnice Observatory. The filled squares denote a part of the light curves
where the amplitude of the brightness changes was typical for HP Lyr i.e.~0.5 
mag in $V$ band. B light curve was shifted by -0.1 mag for clarity of the picture.}
  \label{LiCu}
\end{minipage}
\end{figure}

HP Lyr was one of the targets in our project of $U\!BV\!RI$ photometric 
monitoring of long period binaries in Piwnice Observatory of Nicolaus 
Copernicus University in Toru{\'n} (Poland).
The photometry was carried out during 1998 and 1999 using a one-channel
photometer on the 60-cm Cassegrain
telescope equipped with EMI9558B photomultiplier. 
As a comparison star, we chose a very close to
HP Lyr A0V star HD 182592 ($V$=8.01, $U-B$=0.06, $B-V$=0.07, $V-R$=0.07, $R-I$=0.01). 
The effective wavelengths of our instrumental $ri$ bands were significantly shorter than those 
of Johnson's system: 6390 \AA$\,$ and 7420 \AA, respectively. Nevertheless, for a single star 
we can use the formal transformation, found by observations of Johnson's 
standards: $(R-I)=1.40(r-i)$, and $R=r-0.37(R-I)$.
Our original data are presented on Fig.~\ref{LiCu}. The mean observational errors
were 0.04, 0.03, 0.01, 0.01, 0.03 in particular $UBVri$ bands, respectively. 

\begin{table}                                   
\caption{ Journal of the spectroscopic observations of HP Lyr}\label{Jour}
\tabcolsep2pt                 
\begin{minipage}[bh]{\linewidth}
{\small
\begin{tabular}{|c|c|c|c|c|c|c|c|c|c|c|}
\hline                        
\hline                        
Date  &  JD             & Phase &Exp.& $\lambda\lambda$ &Disp.& Spec.& $T_{eff}$&Rad. vel. & N\footnote{The number of lines} & O\footnote{T -- Tartu Observatory, P -- Piwnice Observatory}\\
& (2\,450\,000+)        &       &(min)&(nm)& (\AA$\!/\!$pix) & class& (K) &(km/s)    &  & \\\hline
26/27.08.00 &   1783.375& 0.693 & 30 &   375-480  & 2.0& A4& 8500& -81 $\!\pm\!$  8    & 3&T \\
05/06.09.00 &   1793.375& 0.765 & 40 &      "       &  " & A5& 8200&-112 $\!\pm\!$ 15    & 3&T\\
09/10.09.00 &   1797.374& 0.794 & 30 &      "       &  " & A5& 8200&-102 $\!\pm\!$ 10    & 3&T \\
17/18.09.00 &   1805.377& 0.851 & 60 &   625-665  & 0.8& &--& -122 $\!\pm\!$ 2         &14&T \\
19/20.10.00 &   1837.258& 0.081 & 60 &      "       &  " & &--& -135 $\!\pm\!$  4        &11&T \\
      "     &   1837.348& 0.082 & 50 &   375-480  & 2.0& A7& 7700&-103 $\!\pm\!$ 12    & 3&T \\
27/28.11.00 &   1876.189& 0.362 & 60 &   625-665  & 0.8& &--& -123 $\!\pm\!$  3        &13&T \\
02/03.05.01 &   2032.433& 0.489 & 20 &   350-550  & 2.0& F2& 7000& -88 $\!\pm\!$ 8     & 5&P \\
08/09.05.01 &   2038.388& 0.529 & 60 &   625-665  & 0.8& &--& -106 $\!\pm\!$ 5         &15&T\\
24/25.05.01 &   2054.538& 0.648 & 10 &   350-550  & 2.0& A3& 8800&-135 $\!\pm\!$ 9     & 5&P \\\hline
\end{tabular}}
\end{minipage}
\end{table}

                              
We have obtained some CCD spectrograms of this star in Tartu Observatory in Estonia 
and two spectra in Piwnice Observatory. The observations were carried out using 
the Cassegrain grating spectrographs attached to 1.5m (Tartu) and 0.9m (Piwnice) telescopes. 
Table~\ref{Jour} presents the journal of all our spectroscopic observations and
results of the radial velocity measurments and the spectral class classification. 
The spectra were reduced under the IRAF and
the MIDAS packages.

\section{Period searching}

\begin{figure}[t]
\begin{minipage}[ht]{0.5\linewidth}
\centering\resizebox{\linewidth}{!}{\includegraphics{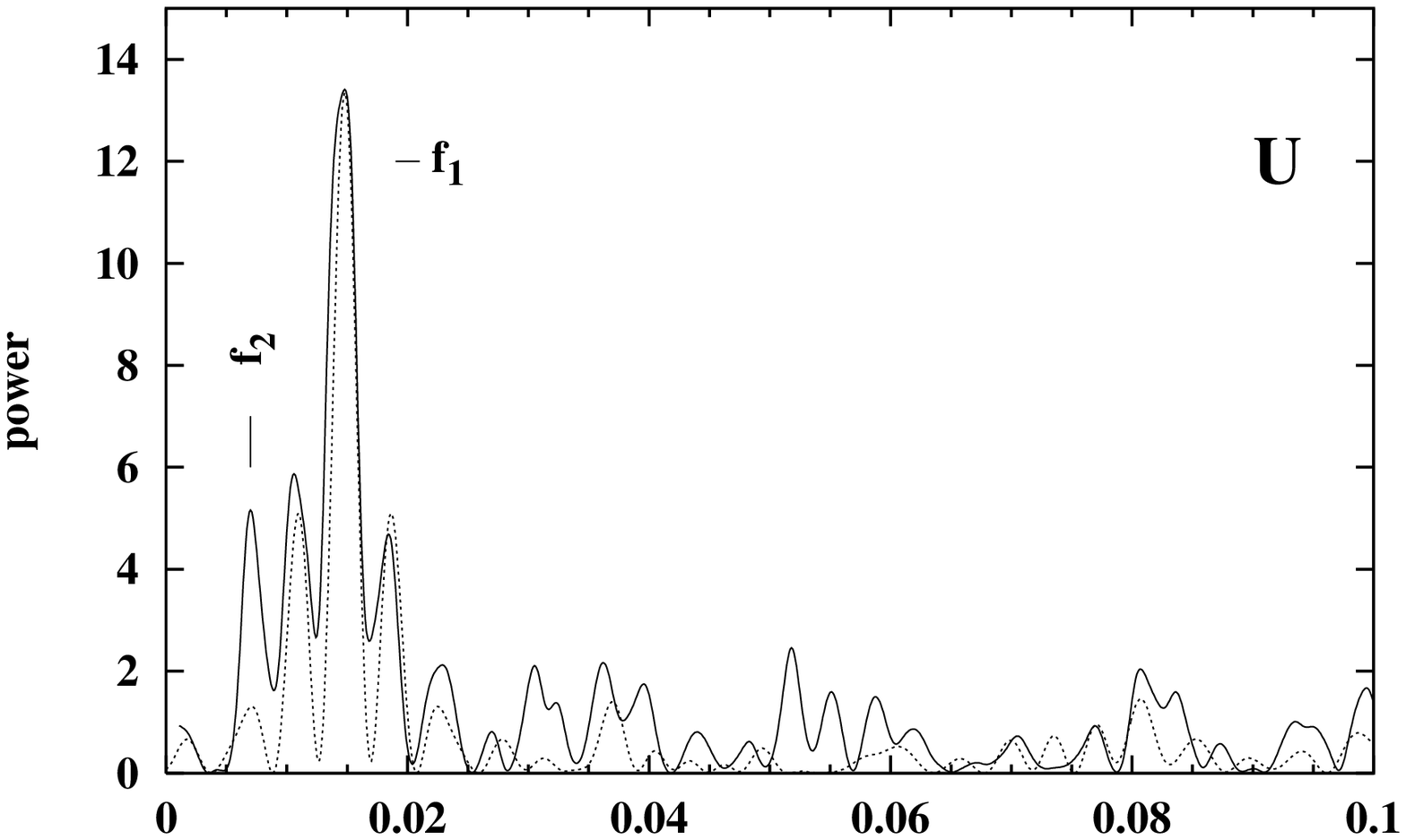}}
\centering\resizebox{\linewidth}{!}{\includegraphics{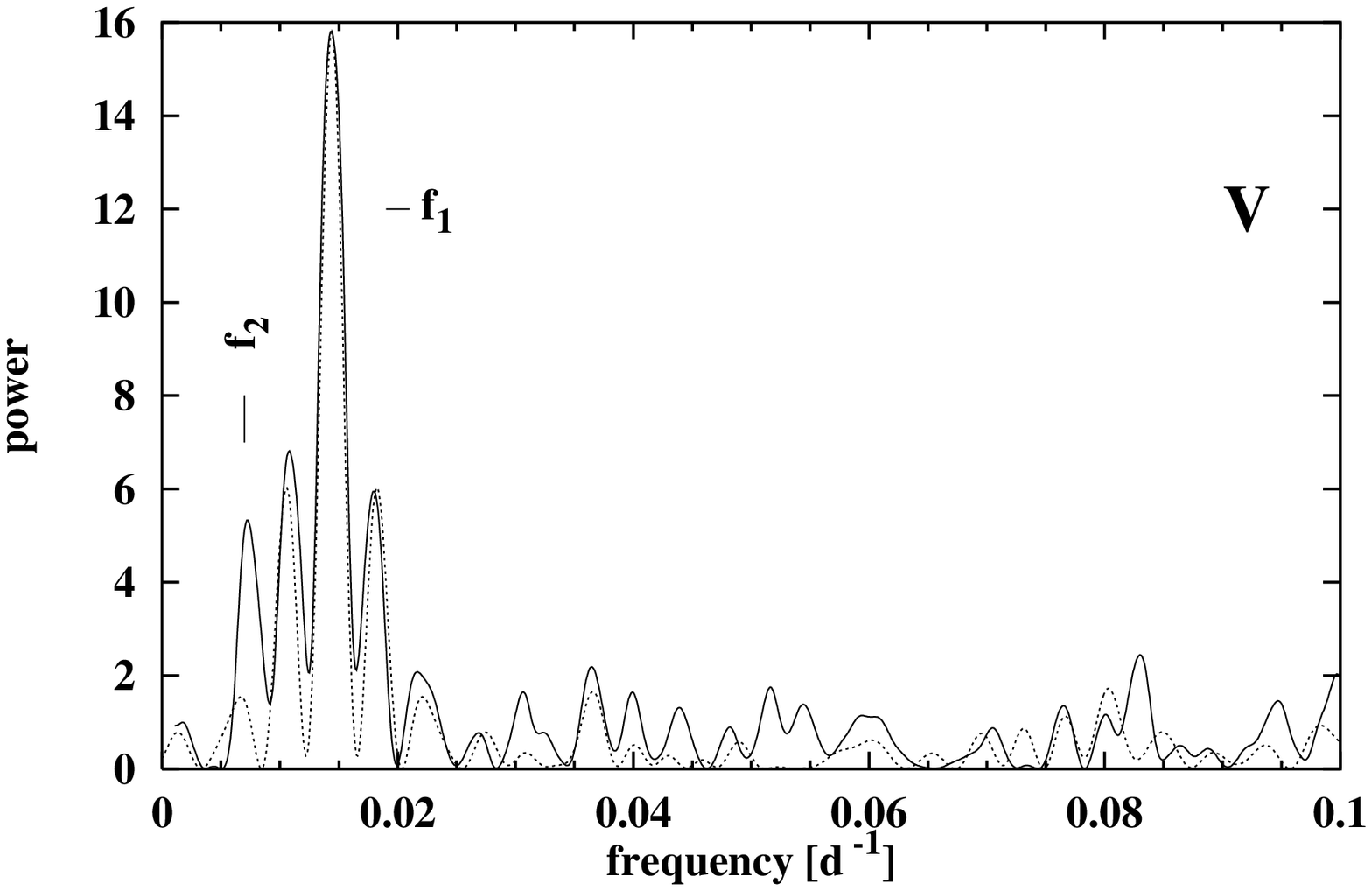}}
\end{minipage}\hfill
\begin{minipage}[h]{0.5\linewidth}
\centering\resizebox{\linewidth}{!}{\includegraphics{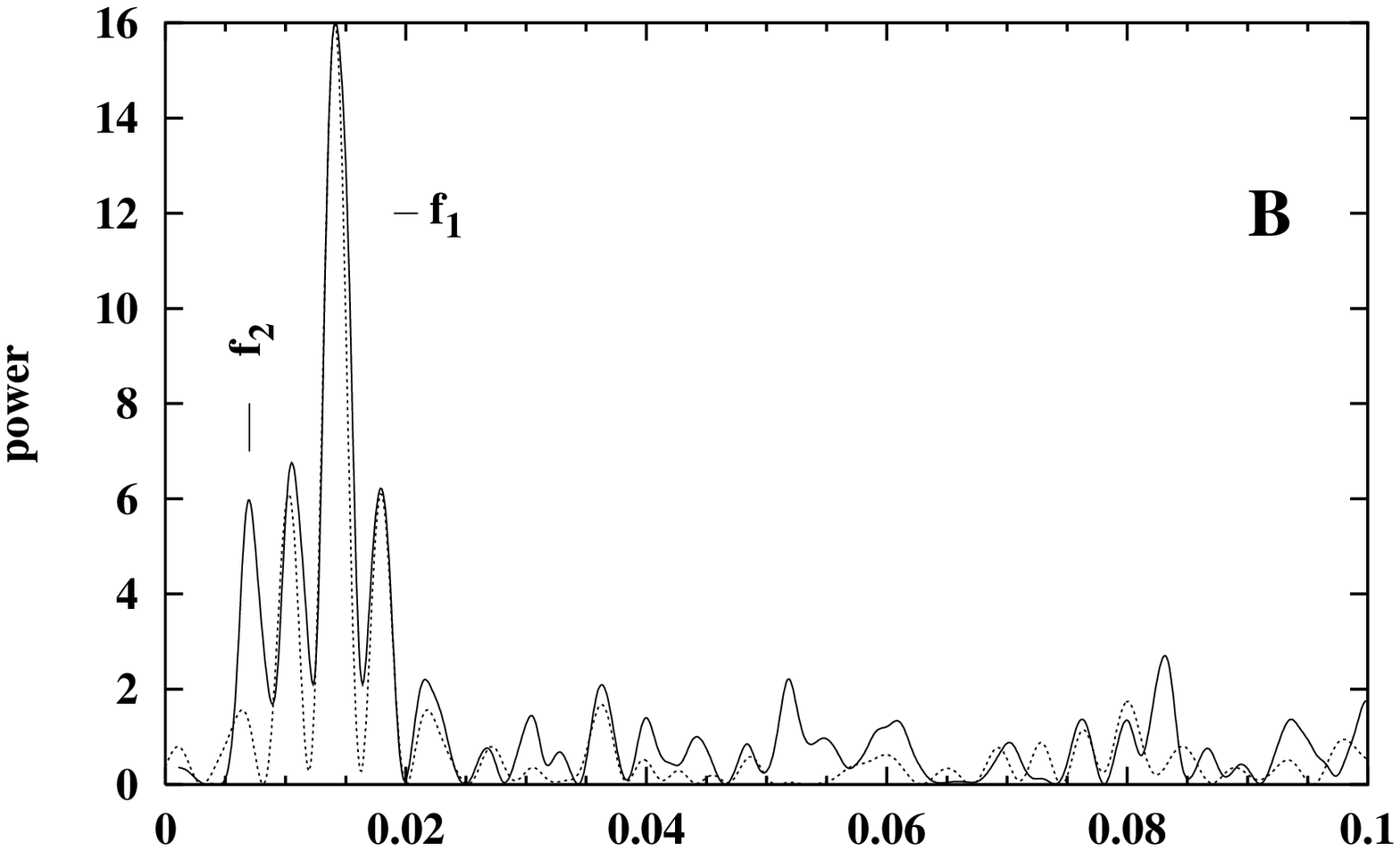}}
\centering\resizebox{\linewidth}{!}{\includegraphics{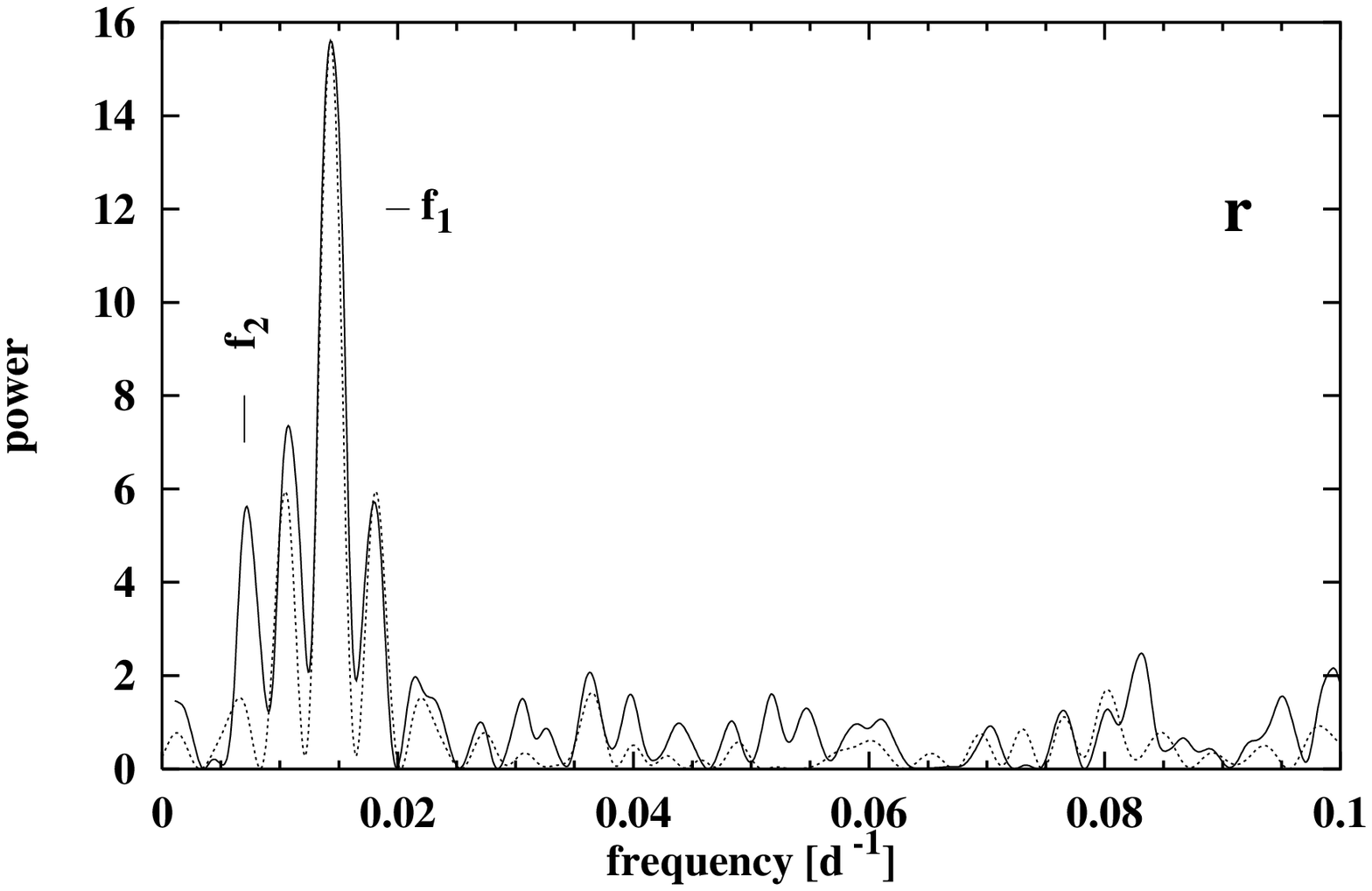}}
\end{minipage}
  \caption{\small The Lomb-Scargle periodograms of the $UBVr$ band observations 
of HP Lyr. The window function is marked with dashed line and centered on the 
highest peak at the frequency $f_1$. 
Note the presence of the signal $f_2$ which
is not an alias of the $f_1$ frequency. 
The omitted diagram for $i$ band has practically the same shape.}
  \label{LoSca}
\end{figure}

     
We have performed frequency analysis of our photometric observations using 
the Lomb-Scargle periodograms (Lomb 1976, Scargle 1982).
In each band we detected a strong peak 
at the frequency of about $f_1=0.0144$ (69.5 days) and the
second one at a frequency $f_2=0.0072$ (139 days), 
whereas two sidelobes around $f_1$ signal reflect the shape of the spectral 
window (Fig.~\ref{LoSca}).
These two detected signals $f_1$ and $f_2$ are in the 2:1 resonance.
The mean period derived from $f_1$ frequency obtained in all filters and assuming 
a 2:1 resonance is $P=139.4 \pm 0.7$ days, i.e. about 1\% less than the previous
estimations (Wenzel 1960, Kreiner et al.~2001).

\begin{figure}
\begin{minipage}{\linewidth}
\centering\resizebox{9.8cm}{!}{\includegraphics{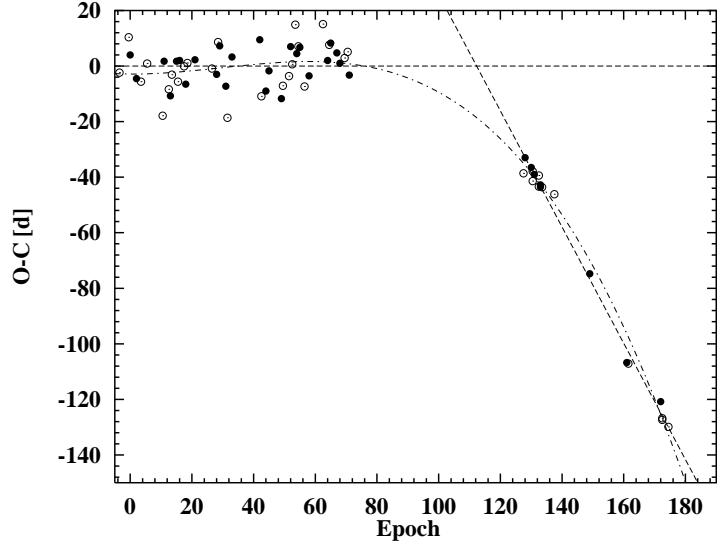}}
\end{minipage}
  \caption{\small O-C diagram for "primary" (dots) and "secondary" (open circles)
minima calculated according to the original Wenzel's ephemeris (Eq.~\ref{Eqn1}).  
Two straight dashed lines correspond to the abrubt period change solution (Eq.~\ref{Eqn1} and~\ref{Eqn2}) and
the dashed-dotted line correspond to the cubic solution: Eq.~\ref{Eph3}.}
  \label{OC1}
\end{figure}

Looking for the ephemeris of HP Lyr we have carried out the timing analysis of observed minima. 
The Wenzel's (1960) original ephemeris:
\begin{equation}
{\rm Min\, I} = {\rm JD}\,\, 2\,426\,910 + 140\fd 75\, E. \label{Eqn1}
\end{equation}
was based on 66 independent moments of 56 photographically observed 
minima between 1931 and 1959. His O-C's residua are shown in Fig.\ref{OC1} for $E<72$.
During 1960-1980 there were no observations mentioned in the literature. 
Since 1981 several moments of minima, estimated visually, were observed by 
Tristram Brellstaff (JD 2444817, 2444893, 2445171, 2445240, 2445510, 2445587, 2446217,
2447807) and published in a number of BAA VSS circulars (Markham and Pickard 2001).
The next set of minima dates was collected by Kreiner et al.~(2001). These data contain 
photoelectric and visual observations done by W.~Braune et 
al.~(JD 2445236.5, 2445309.2) and J.~Heubscher et al.~(JD 2445516, 2445586, 2445656.5, 
2449464.0, 244953.0, 2450998.2, 2451062.6) and were published in several issues of B.A.V.Mitt.   
We have also added two moments of minima from our observations: JD 2425062.0 and JD 2451341.0 (Fig.~\ref{LiCu}).

All data after 1980 ($E>120$) show significant deviation from Wenzel's ephemeris (Fig.~\ref{OC1}).
We found a satisfactory, cubic solution connecting all minima which is presented in Fig.~\ref{OC1}: 
\begin{equation}
{\rm Min\, I}  = {\rm JD}\,\, 2\,426\,907 + 140\fd 74\, E 
 - 0\fd 0043\, E^2 - 4\fd 9\!\cdot\! 10^{-5}\, E^3 \label{Eph3}
\end{equation}
However the second possibility -- the linear equation for the later data -- gives a slightly better 
fit assuming the existence of an abrupt period decrease:
\begin{equation}
{\rm Min\, I} = {\rm JD}\,\,  2\,444\,893 + 138\fd 66\, (E-128). \label{Eqn2}
\end{equation}
The numbers $E$ are the same as in Wenzel's ephemeris (Eq.~\ref{Eqn1}). Epoch $E=128$ corresponds
to the first observed "primary" minimum after the 1960-1980 gap. If the abrupt period
change is real, it should most probably occur in 1974 at the 112th cycle according to Wenzel's ephemeris.  
Observations should soon distinguish between the cubic and the linear ephemeris.

\section{Eclipsing or pulsating star?}
\begin{figure}
\begin{minipage}{\linewidth}
\centering\resizebox{0.63\linewidth}{!}{\includegraphics{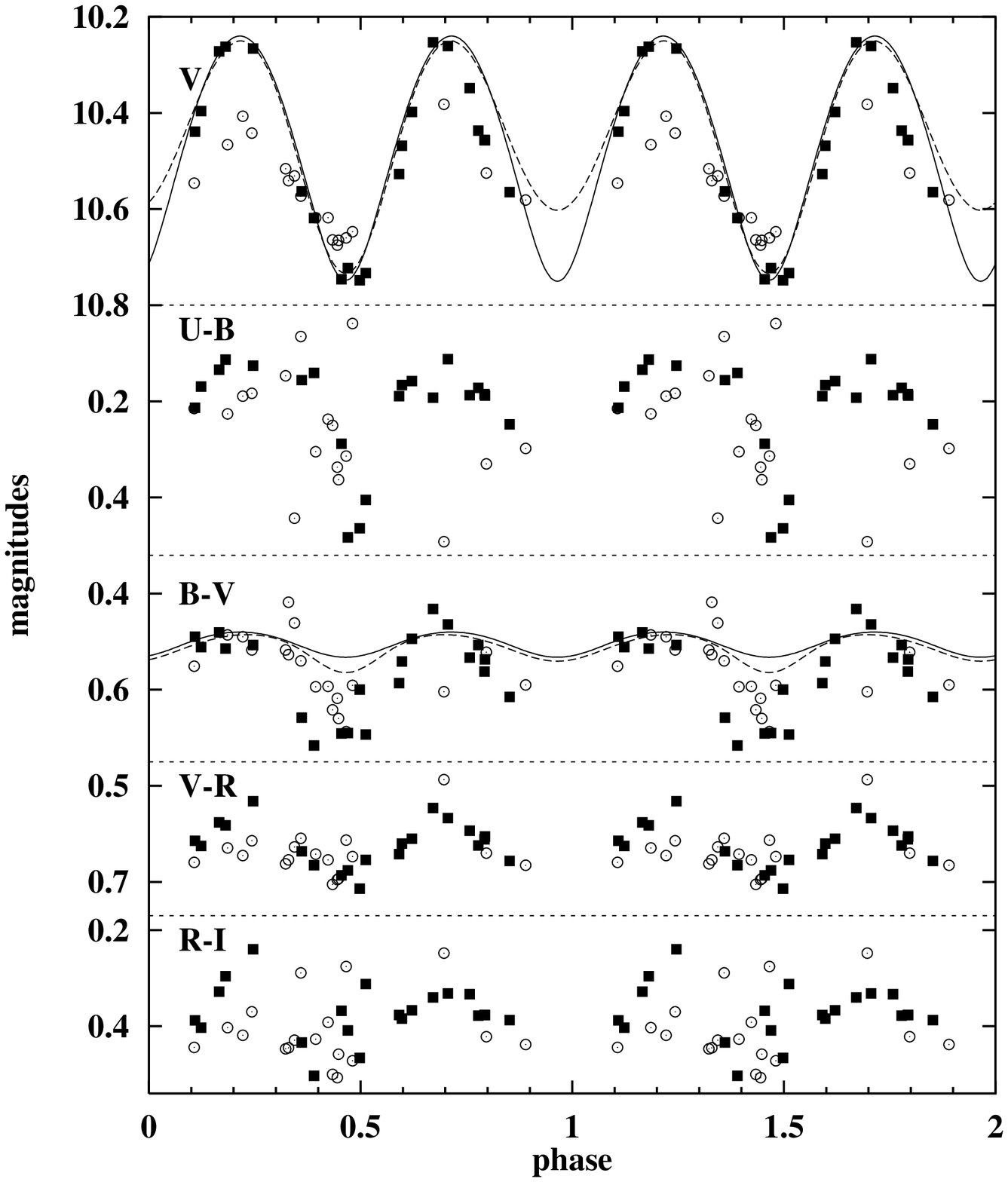}}
  \caption{\small $V$ photometry and colour indices folded up according to Eq.~(\ref{Eqn2}).
  The fits with the extremal gravitation darkening effect are shown with $V$ and $B-V$ data.
  Continuous line: overcontact model for two A6 giants ($\Omega_1$=$\Omega_2$=3.27, $i$=62$^o$), 
  dashed one: semidetached model with invisible companion ($q$=M$_{\rm A6}$/M$_{inv}$=0.2, $i$=90$^o$).}
  \label{NLiC}
  \end{minipage}
  \mbox\\
\begin{minipage}{\linewidth}
\centering\psfig{file=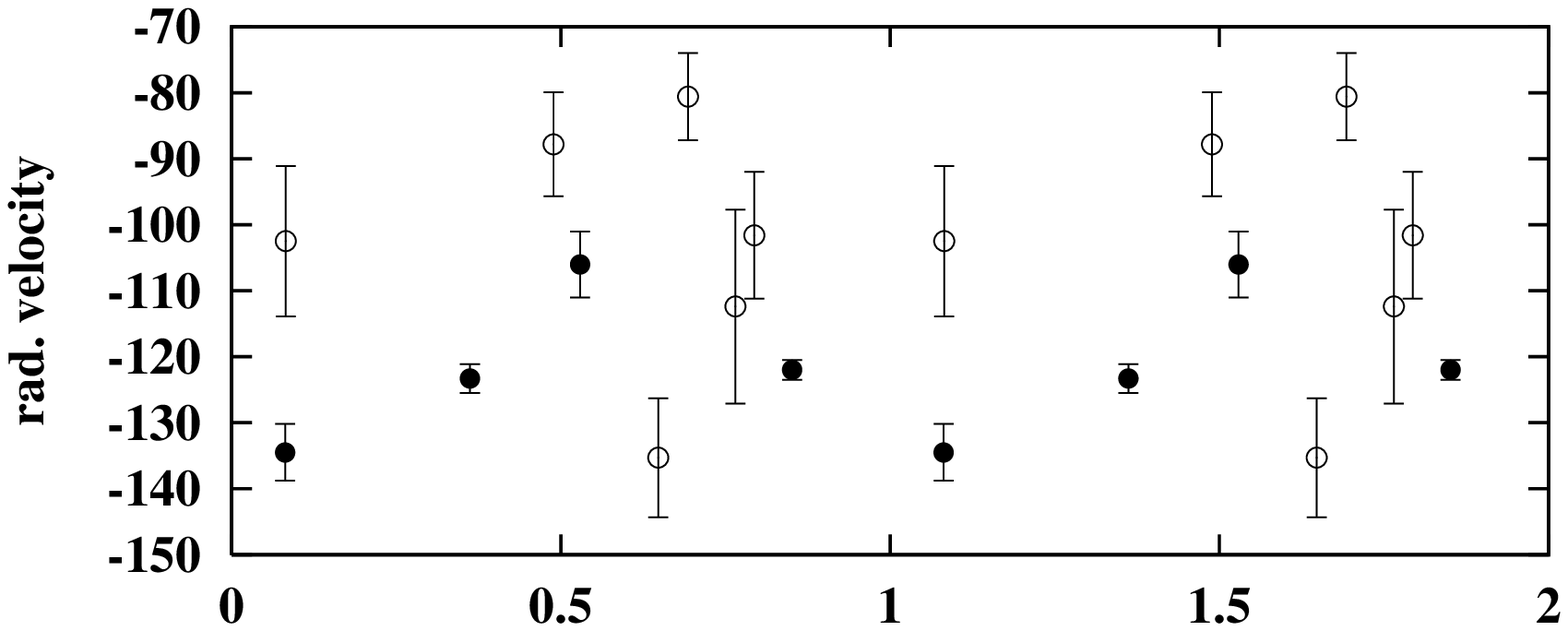,width=0.62\linewidth}
\centering\psfig{file=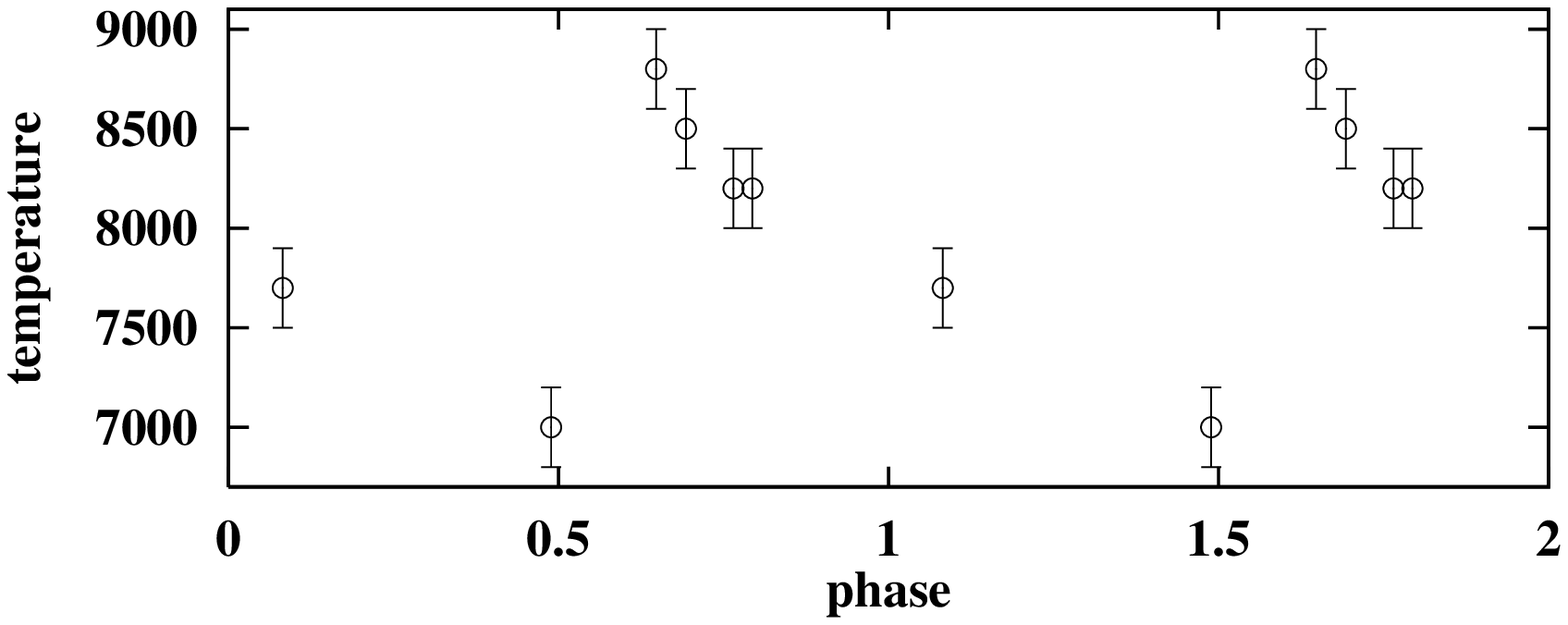,width=0.62\linewidth}
\caption{\small {\it Upper panel:} radial velocities of HP Lyr from Balmer lines 
(blue spectra -- open circles) and metallic lines (red spectra -- filled circles). {\it Bottom:} the effective 
temperature resulting from derived spectral type.   \label{radtem}}
\end{minipage}
\end{figure}

Our photometry shows that initially the amplitude of the light variations was about 0\fm 5 in $V$ band,
slightly smaller in red $ri$ filters and was raising in blue filters up to 1\fm 0 ($U$ band). 
But, after about JD 2451150, all amplitudes decreased by a factor of two, whereas the mean
brightness remained unchanged (Fig.~\ref{LiCu}). This is rather not typical behaviour 
for an eclipsing binary and we decided to show the mean light curves using $V-R$ and
$R-I$ transformed Johnson colours (Fig.~\ref{NLiC}).
Our $V$ light curve in Fig.~\ref{NLiC} resembles the $\beta$ Lyr type curve with 
a slightly deeper "primary" minimum at phase 0.5. However, we have a gap in observations
around phase 1.0. Also, the colour 
index curves seem to be deeper at phase 0.5. This fact is confirmed by our
spectral observations which give earlier spectral type and higher temperature around phase 1.0 than
around phase 0.5 (Fig.~\ref{radtem}). Wenzel (1960) reported that both minima
are of similar depth but the inspection of his photographic light curve shows
that the minimum numbered by him as 1 (phase 0.0) may be slightly deeper. This alternation between the depth of minima
is very difficult to understand in a binary system. It is real if the cubic solution 
(Eq.~\ref{Eph3}) is a valid model of both sets of the data
in Fig.~\ref{OC1}. On the other hand, if the binary consists of two similar
stars, then any mass loss or exchange of matter should lead to an increase (not a decrease) of the orbital period.

\begin{figure}
\begin{minipage}{\linewidth}
\centering\psfig{figure=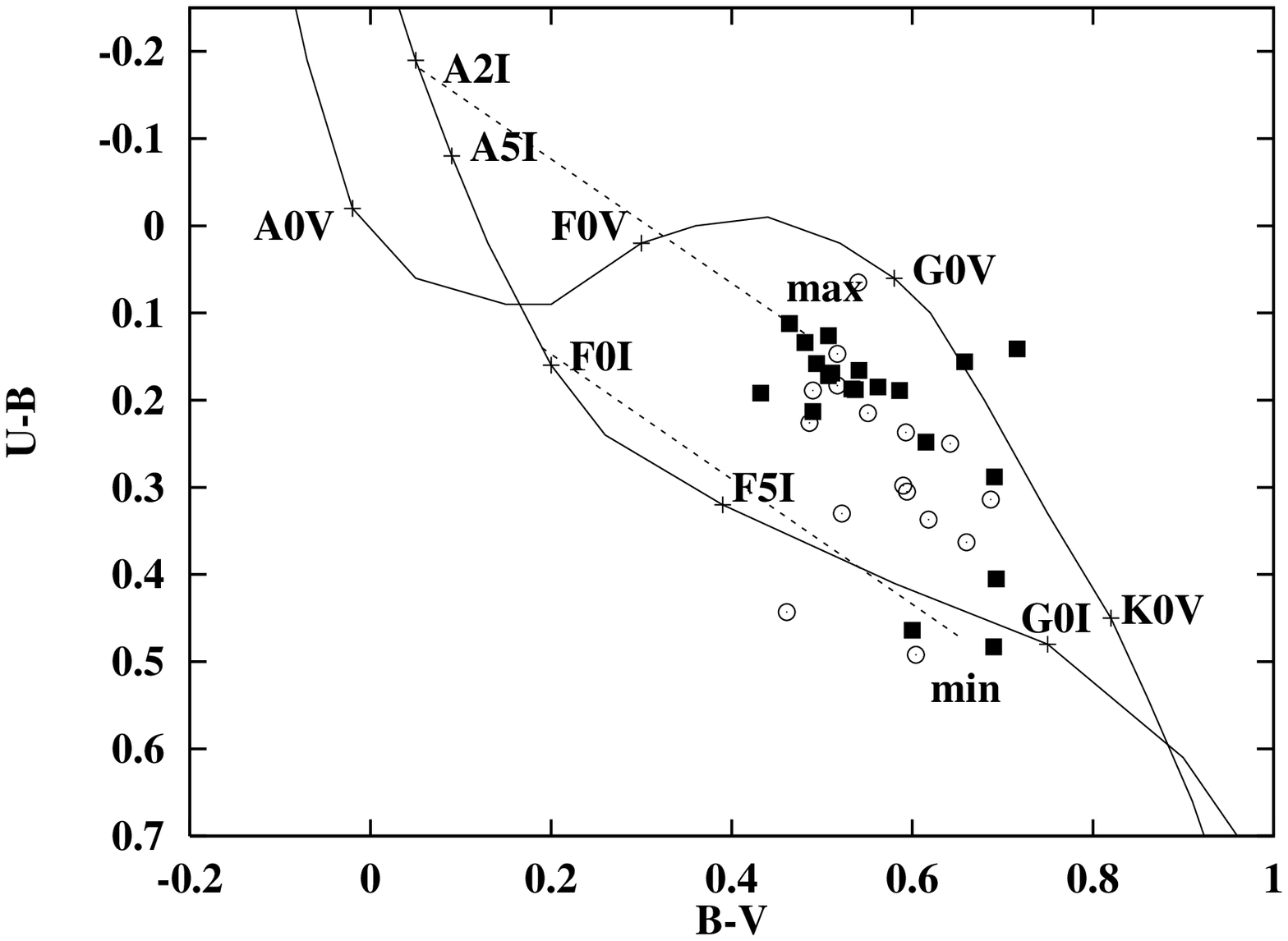,width=0.64\linewidth}
\end{minipage}
  \caption{\small $U-B$, $B-V$ colour-colour diagram of our photometry of HP Lyr. 
Main sequence and supergiant theoretical colours are marked by continuous lines
(Straizys 1977). Dashed lines -- the reddenings of the maxima and the minima 
correspond to $E(B-V)\sim 0.42$. Filled squares and open cirles have the same meaning
as in Fig.~\ref{LiCu}.} 
  \label{col}
\end{figure}

In order to rule out definitively the binary hypothesis we have calculated synthetic light curves using
WD code (Wilson \& Devinney 1971). In general, only the ellipsoidal variations in a binary system can produce reddening of both minima.
Especially in early type stars with radiative envelopes, a significant gravitational
reddening effect should be expected following von Zeipel's (1924) theorem. We have tested two possible models with extreme von Zeipel's 
effect: 1) an overcontact binary with two similar and evolved A6 components 
and 2) a semidetached binary consisting of an A6 star filling its Roche-lobe and massive, compact,
optically invisible companion. Both ellipsoidal models can roughly reproduce the V band light curve but 
failed to reproduce the colour variations -- Fig.~\ref{NLiC}. No binary model can explain the observed 
reddening with the amplitude about $\Delta (B-V)\approx 0.3$ during both minima.

\begin{figure}
\begin{minipage}[h]{\linewidth}
\centering\resizebox{\linewidth}{!}{\includegraphics{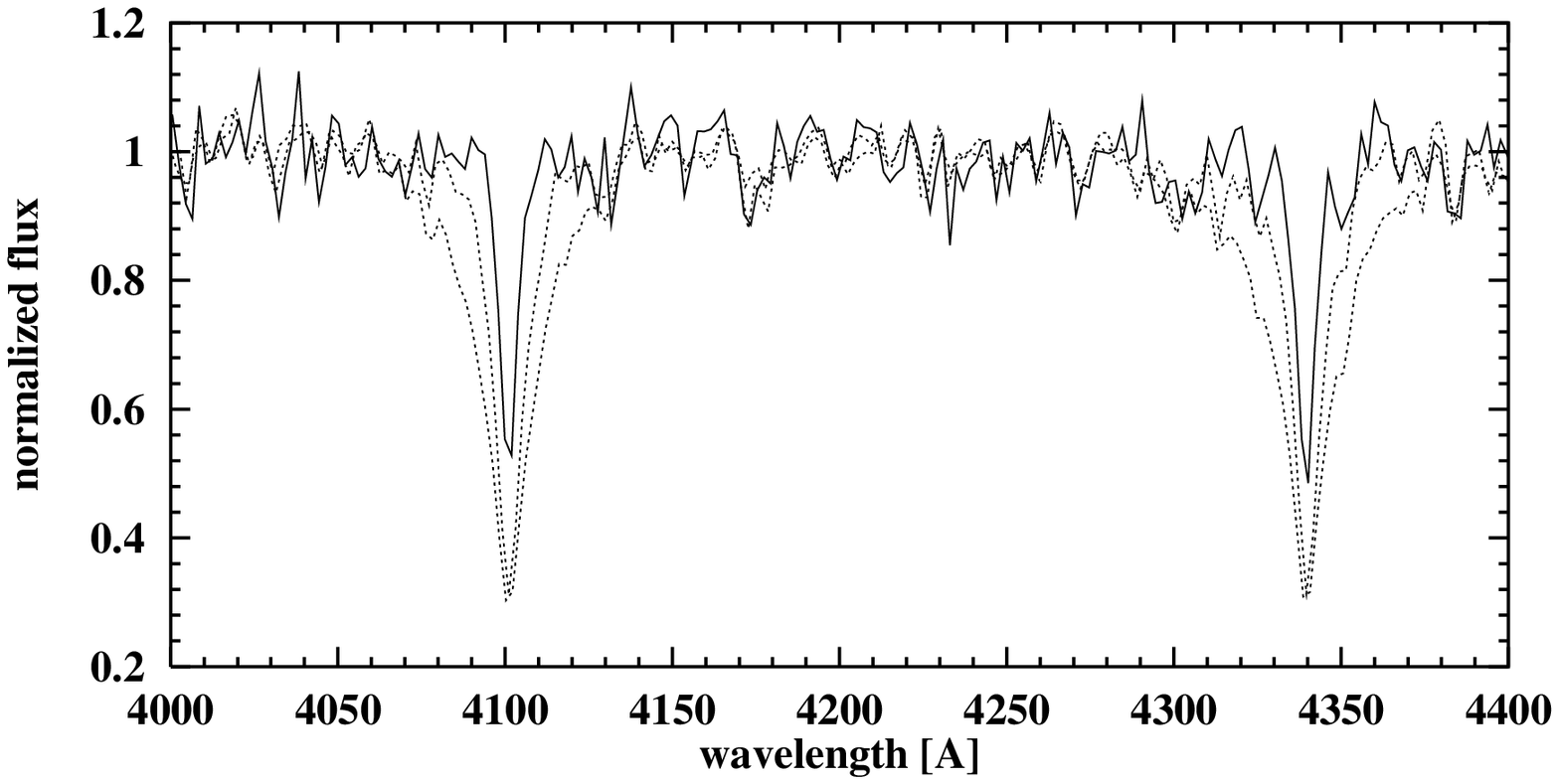}}
  \caption{\small The luminosity sequence of A-type stars in profiles of H$\delta$ and H$\gamma$ lines.
  Dotted lines: HD 178187 (A4III) - the broadest absorption profiles  and HD 186177 (A5Ib).
The spectrum of HP Lyr -- continuous line -- was taken on 24/25 May 2001 (at maximum) and indicates
A3 Ia-Iab class. Note sharp Balmer lines and intensive Fe II and Ti II blend 
at 4172-8 \AA~and Fe II 4233 line in the HP Lyr spectrum.} 
  \label{Lumclas}
\end{minipage}   
\begin{minipage}[h]{\linewidth}
\centering\resizebox{\linewidth}{!}{\includegraphics{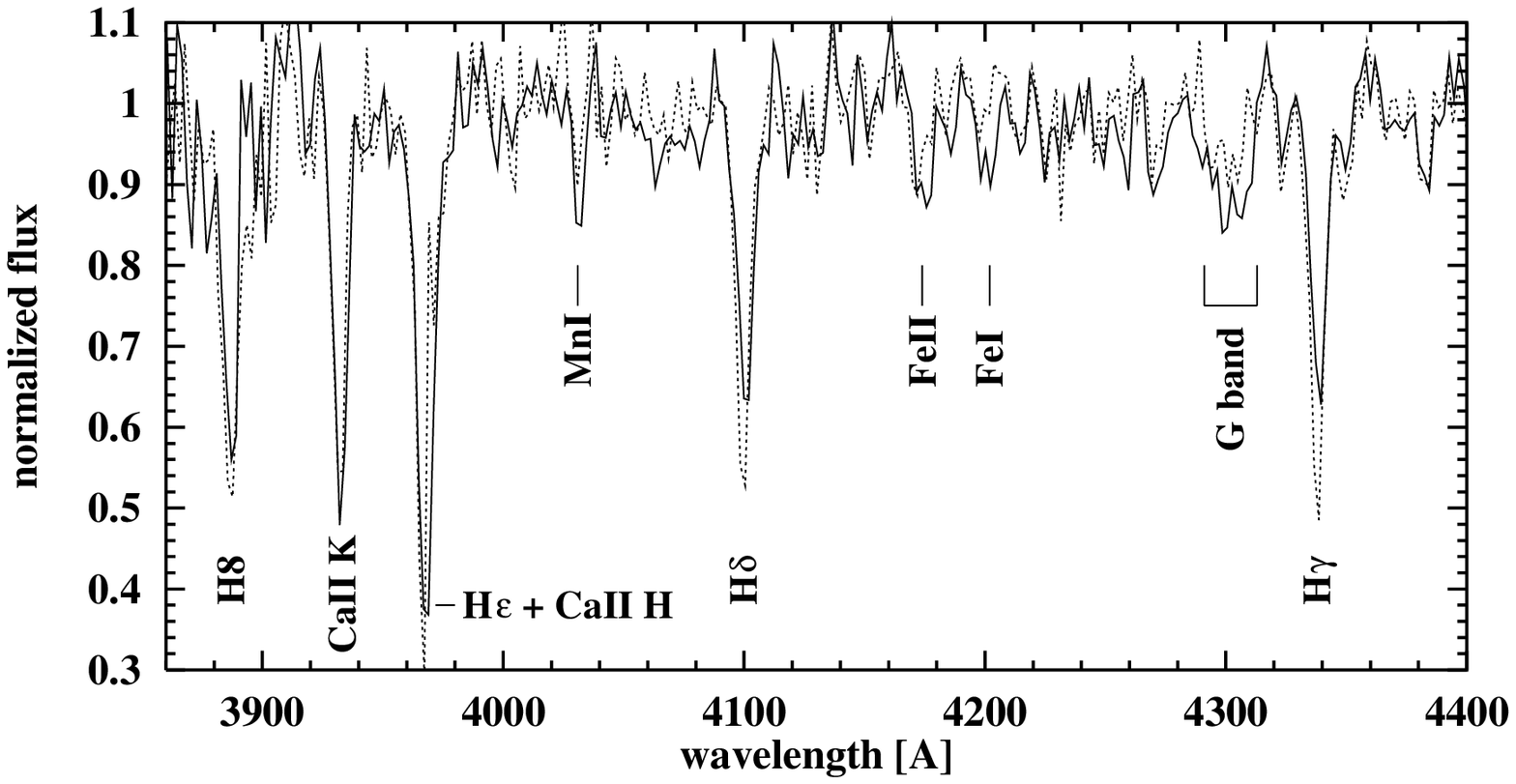}}
  \caption{\small The comparison between two spectra, obtained on 2/3 May 2001  
   exactly at the minimum (F2 type, solid line) and 24/25 May 2001 at the maximum 
   (A3 type dotted line). Note the increasing of CaII and metallic lines, 
   together with the appearance of the G-band in F2 spectrum,
   and significant increasing of Balmer absorption in A3 spectrum.}
  \label{BluC}
\end{minipage}
\end{figure}

Changes of colours create complex loops on the $U-B$, $B-V$ diagram (Fig.~\ref{col}).
However, they show a general trend to align with the supergiant sequence from A2 at maximum
to F0 at minimum with constant $E(B-V)=0.42$. 
Nearly the same spectral changes from A3 at maximum to F2 at minimum (Fig.~\ref{BluC}), were obtained
from our blue spectra (Table~\ref{Jour}) using MK criteria by Morgan et al.~(1978).
The temperature changes  derived from Straizys' (1982) calibration
strictly follow the light and colour variations (Fig.~\ref{radtem}). 
The most probable explanation of this behaviour is pulsations of
a single star.
Although our radial velocity data are insufficient for interpretation in 
terms of pulsations, they can preclude the binary model. 
It is not possible to join the four radial
velocity points obtained from metallic lines (Fig.~\ref{radtem}) by one sinusoidal line
with the condition of crossing the barycentric velocity close to
spectroscopic conjunctions ("eclipses" at phases 0.0 and 0.5).

\begin{figure}
\begin{minipage}[th]{\linewidth}
\centering\psfig{figure=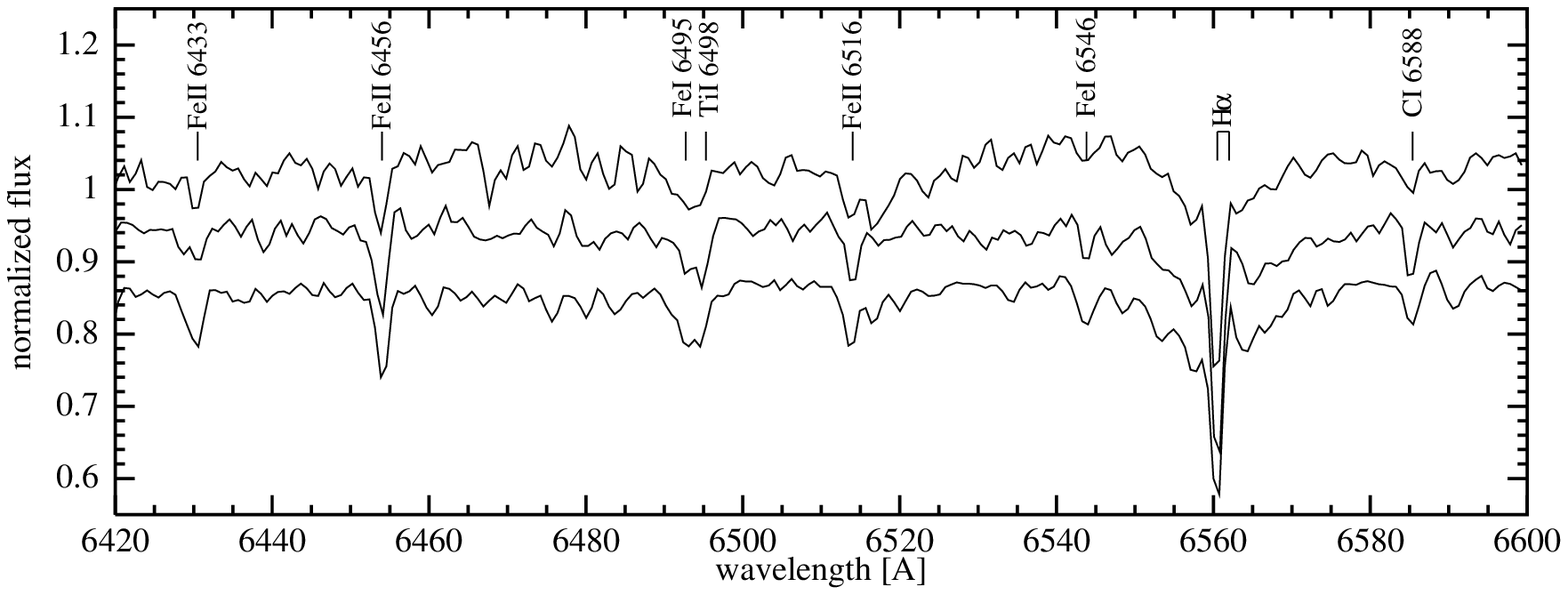,width=\linewidth}
\caption{\small Comparison of the 6420 - 6600 \AA~region 
   of HP Lyr's spectrum around H$\alpha$ line obtained on 19/20 October 2000 (uppermost), 
   27/28 November 2000 and 17/18 September 2000 (lowest).  All the spectra were taken in Tartu at 
   phase 0.08, 0.35 and 0.85, respectively.  \label{Hacom}} 
\end{minipage}
\end{figure}

\section{Evolutionary status}
The supergiant luminosity class Ib-Iab was derived by comparison with 
the spectra of the MK standards -- Fig.~\ref{Lumclas}. 
However, relatively high galactic latitude 
($b=+11^{o}\!\!.7$) and high radial velocity ($-$113 km/s) indicates 
that HP Lyr is most likely an evolved star of Intermediate Population II. 
The massive young object of Population I stars should be expected close
to the Galactic plane. The spatial distribution of the interstellar extinction
at the same galactic longitude ($l=70\fg - 75\fg$), but lower latitude ($b=5\fg - 6\fg$) shows
practically constant extinction $A_v = 1.2-1.6$ above a distance of about 1.5 kpc (Neckel \& Klare 1980, 
Miko{\l}ajewska \& Miko{\l}ajewski 1980).
This value of $A_v$ corresponds well to $E(B-V)$ for HP Lyr and gives 
a limit for the absolute magnitude $M_v < -$1\fm 5.

A comparison between the spectra in the $H_{\alpha}$ region 6420 - 6600 \AA \,obtained 
during two descending branches of the light curve is 
presented in Fig.~\ref{Hacom}. 
There are numerous metallic absorption lines used for radial velocity measurements (Table~\ref{Jour}). 
On the blue spectra, only three Balmer lines (and additionally CaII K and FeII 5169 \AA$\,$
at Piwnice) were measured. 
The mean heliocentric radial velocity measured from the Balmer 
lines is  $-104\pm 5$ km/s and from metallic lines $-122\pm 5$ km/s. 
It seems that all extremaly sharp Balmer absorptions are affected by the P Cyg emission components related to
shock expanding in the atmosphere. Such weak emmission 
is clearly visible in the $H_\alpha$ profile in Fig.~\ref{Hacom}.

HP Lyr was positionally associated with IRAS source 19199+3950 by
Friedemann et al.~(1996), but they rejected this identification 
because of its early spectral type. 
HP Lyr has the very good positional coincidence $3^"$ with an IRAS source 
(it is much better then most identified
IRAS sources in their catalog), whereas the weak red star which Friedemann 
et al. suggested as a possible optical counterpart lies 8 times further 
(i.e. at the distance $24^"$). 
We found HP Lyr in maps of the Two Micron All Sky Survey\footnote{\mbox{The 
2MASS Internet Archive is avalaible at the webpage 
{\it http:{//}www.ipca.caltech.edu{/}2mass}}} with $J=8.98$, $H=8.44$ and $K=7.73$.
Within a radius of 60$^{"}$ around HP Lyr there is no other $JHK$ source. 
We conclude that HP Lyr is the only countepart of the IRAS infrared source.
Such infrared excess is typical for evolved Population II stars.

\section{HP Lyr as the RV Tau variable}
 
The $\beta$ Lyr type shape of the light curve and the other observed
photometric and spectroscopic properties
of HP Lyr show many important similarities with RV Tau group of variables.
The RV Tau stars are luminous, pulsating variables located 
in the brightest part of the Population II instability strip and overlap 
in the Hertzsprung-Russell  diagram 
with the W Vir type II Cepheids (Wahlgren 1992). 
Typical members have spectral type between F and K, 
luminosity class Ia-II, periods of their light variations in the range
from about 30 to 150 days, large spectral and colour changes from maxima 
to minima. Alternating deep and shallow minima are caused by two dominant frequencies in their power spectra 
in 2:1 resonance 
with the ratio of their amplitudes close or smaller than unity.  
There is general agreement that RV Tau stars are pulsating low-mass post-AGB
stars in the transition into the planetary nebulae phase (Jura 1986, 
Giridhar et al.~2000). 

Most photometric peculiarities of HP Lyr are also typical for RV Tau stars.
The photometric study of Pollard et al.~(1996) revealed several objects
such as AR Pup which during the period of observations 
did not show any distinction between primary and secondary minima (just like HP Lyr).
Since the mean brightness of HP Lyr remains unchanged, it belongs to the RVa subgroup of RV Tau stars.
Possible alternation of the "primary" and the "secondary" minima observed in HP Lyr
is also typical for RV Tau stars (e.g.~R Sct). The
asymmetric, "bowed" shape of the colour curves are typical for these variables.

Another spectacular feature observed in HP Lyr photometric behaviour was
the period change(s). The pulsation period of the RV Tau stars often varies with a complex 
way in a long timescale. Their O$-$C diagrams show many period instabilities 
and abrupt changes when the period increses or decreases from 0.001$P$ to 0.01$P$
(Erleksova 1971, Percy et al.~1991). The timescale over which the period changes can be from 20
to over 100 cycles. 

The observed spectroscopic features of HP Lyr correspond quite well 
to common spectroscopic characteristics of RV Tau stars (Pollard et 
al.~1997 and references therein). 
For example, from many $H_\alpha$
profiles of some RV Tau stars, IW Car profiles are most similar
to those of HP Lyr.  
IW Car itself was reported to be a binary system with spectral 
classification of A4 Ib-II: + F7/8 (Houk 1987), but the radial 
velocity measurements  contradict this hypothesis.
Most of the RV Tau type stars have more prominent P Cyg profiles in the $H_\alpha$ lines. 
It can be explained by their lower temperature and weaker photospheric components of
the Balmer absorptions.
One of our spectra of HP Lyr taken at a
minimum (2/3 May) shows a weak CN I blend at 3883. Thus, in the spectroscopic 
subclassification scheme proposed by Preston et 
al.~(1963), HP Lyr should belong to the
the spectroscopic group "B" as an extremely hot pulsating RV Tau star (mean $T_{eff}\sim 7700$ K). 
The hottest RV Tau star known previously, IW Car, has the temperature $T_{eff}=6700$ (Giridhar et al.~1994).

\begin{figure}
\begin{minipage}{\linewidth}
\centering\psfig{figure=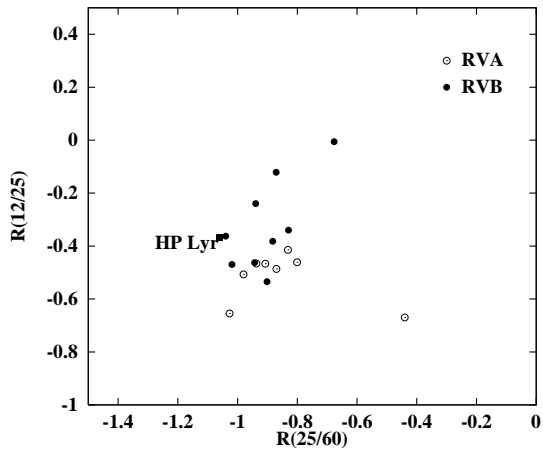,width=7.3cm}
  \caption{\small R(12/25) -- R(25/60) colour diagram of the RV Tau stars detected
by IRAS. The data taken from Raveendran (1989). The position of 
HP Lyr was marked by filled square. \label{Iras}}
\end{minipage}
\end{figure}

Many RV Tau stars also exhibit
infrared excesses (Raveendran 1989) from, most probably, 
dusty circumstellar shells.
The presence of infrared excess in the case of RV Tau stars may be a sign
of their old evolutionary stage as AGB stars. On the R(12/25) -- R(25/60)
diagram HP Lyr lies in the area populated by RVB stars (Fig.~\ref{Iras}).    
Also, the colours: $J-H = 0.54$ and $H-[25] = H+2.5\log(F_{25\mu m}) = 9.88$ of HP Lyr 
place it inside the well defined area populated by RV Tau stars
at the near- far-infrared colour-colour diagram (Fujii et al.~2001)

Alcock et al.~(1998) derived a single P-L relation for
blue RV Tau stars in the Large Magellanic Cloud (Eq.~6 in their paper). 
If we apply this relation
for the halfperiod of 70 days, we estimate the absolute magnitude of HP Lyr: $M_v \sim -4.5$. 
This value implies the distance to the star $\sim$ 5 kpc and the distance from the galactic plane $z\sim 1$ kpc.
The high absolute magnitude of HP Lyr implied by Alcock et al.'s~P-L relation may not
be correct. However,           
the low amplitude of brightness changes in HP Lyr
($\Delta V = 0.5$ mag) is in a very good agreement with the slope of the relation $V$ amplitude-period
in Fig.~7(b) of Alcock et al.(1998)

\section{Conclusions}
The results of the photometric and spectroscopic survey of the long period variable
HP Lyr have been reported. 
The star has behaviour typical for RV Tau stars but is apparently hotter
than any other known RV Tau object. It means that this star makes a substantial extension
to the Type II Cepheids/RV Tau instability strip to higher temperatures. 
Thus HP Lyr can be very important from
an evolutionary point of view. 
Further photometric monitoring is very interesting to follow the period
changes.

{\bf Acknowledgements.} 
We are thankful to J.L.~Janowski for making part of the photometric
observations and to T. Eenm\"ae for taking one spectrum of HP Lyr. We thank
P.~Moskalik for very fruitful comments in the beginning of this work.
We thank J.M.~Kreiner and
T.~Markham and R.~Pickard for kindly making their times of HP Lyr minima data available. 
We are grateful to U.~Maciejewska and B.~Roukema for English corrections in the text.
This study was supported by Polish KBN Grant No.~5 P03D 003 20 and Estonian
Science Foundation grants No.~3166 and 5006.\\[0.1cm]

\begin{center}
REFERENCES\\
\end{center}

\noindent
{Alcock, C. et al. (MACHO collaboration)}, {1998}, {AJ}, {115}, {1921}\\
{Erleksova, G.E.}, {1971}, {Peremennye Zvezdy}, {18}, {53}  \\
{Friedemann, C., Guertler, G., Loewe, M.}, {1996}, {A\&AS}, {117}, {205}  \\
{Fujii, T., Nakada, Y., Parthasarathy, M.}, {2001}, {in "Post-AGB objects \mbox{as a phase} \mbox{\hspace*{0.3cm}} of stellar evolution", Ed.~R.~Szczerba and S.K.~G{\'o}rny, ASSL 265, Kluwer},{~}{111}\\
{Giridhar, S., Lambert D., Gonzales G.}, {2000}, {ApJ}, {531}, {521} \\
{Giridhar, S., Rao, K.N., Lambert, D.}, {1994}, {ApJ}, {437}, {476} \\ 
{Houk, N.}, {1978}, {Michigan Spectral Catalogue of Two-dimensional Spectral Types \mbox{\hspace*{0.3cm}} for the HD
  Stars, University of Michigan, Ann Arbor}, { Vol.~1}{~} \\ 
{Jura, M.}, {1986}, {ApJ}, {309}, {732} \\ 
{Kreiner, J.M., Kim, C., Nha, I.}, {2001}, {"O$-$C Diagrams of Eclipsing \mbox{Binaries", Wy-} \mbox{\hspace*{0.3cm}}dawnictwo Naukowe AP, Krak{\'o}w}{~}{~}   \\ 
{Lomb, N.R.}, {1976}, {Ap\&SS}, {39}, {447} \\ 
{Markham, T., Pickard, R.}, {2001}, {private communication}{~}{~}\\
{Miko{\l}ajewska J., Miko{\l}ajewski M.}, {1980}, {AcA}, {30}, {347}\\
{Morgan, W.M., Abt, H.A., Tapscott, J.W.}, {1978}, {"Revised MK spectral \mbox{atlas for} \mbox{\hspace*{0.3cm}} stars earlier than the sun", Yerkes Observatory}, {~}, {~}\\
{Morgenroth, O.}, {1935}, {Astron.~Nachr.}, {255}, {425}  \\ 
{Neckel, Th., Klare, G.}, {1980}, {A\&AS}, {42}, {251}\\
{Payne-Gaposchkin, C., Brenton, V.K., Gaposchkin, S.}, {1943}, \mbox{{Harvard Ann.}, {113}, {1}}  \\ 
{Percy, J.R., Sasselov, D.D., Alfred, A., Scott, G.}, {1991}, {ApJ}, {375}, {691}  \\ 
{Pollard, K.R., Cottrell, P.L., Kilmartin, P.M., Gilmore, A.C.}, {1996}, \mbox{{MNRAS}, {279}}, \mbox{\hspace*{0.3cm}} {949}  \\  
{Pollard, K.R., Cottrell, P.L., Lawson, W.A., Albrow, M.D., Tobin, W.}, {1997}, \mbox{\hspace*{0.15cm}} \mbox{\hspace*{0.3cm}} {MNRAS}, {286}, {1}  \\ 
{Preston, G.W., Krzemi{\'n}ski, W., Smak, J., Williams, J.A.}, {1963}, {ApJ}, {137}, {401} \\ 
{Raveendran, A.V.}, {1989}, {MNRAS}, {238}, {945} \\ 
{Sandig, H.-U.}, {1939}, {Astron.~Nachr.}, {276}, {177} \\ 
{Scargle, J.D.}, {1982}, {ApJ}, {263}, {835}\\ 
{Straizys, V.}, {1977}, {"Multicolour Stellar Photometry", Mokslas, Vilnius}{~}{~}  \\ 
{Straizys, V.}, {1982}, {"Metal-deficient Stars", Mokslas, Vilnius}{~}{~}  \\ 
{Van der Veen, W.E.C.J., Habing, H.J.}, {1988}, {A\&A}, {194}, {125}  \\ 
{von Zeipel, H.}, {1924}, {MNRAS}, {84}, {665} \\ 
{Wahlgren, G.M.}, {1992}, {AJ}, {104}, {1174}  \\ 
{Wenzel, W.}, {1960}, {Mitt.~Ver.~Sterne}, {499-500}{~}  \\ 
{Wilson, R.E., Devinney, E.J.}, {1971}, {ApJ}, {166}, {605}  \\ 
\end{document}